\documentclass[12pt]{article}

\usepackage{amssymb}
\let\a=\alpha

\newcommand{\beq}{\begin{equation}}
\newcommand{\eeq}{\end{equation}}
\newcommand{\beqn}{\begin{eqnarray}}
\newcommand{\eeqn}{\end{eqnarray}}

\newcommand{\eps}{\epsilon}

\newcommand{\be}{\begin{equation}}
\newcommand{\ee}{\end{equation}}
\newcommand{\ba}{\begin{eqnarray}}
\newcommand{\ea}{\end{eqnarray}}
\newcommand{\bdm}{\begin{displaymath}}
\newcommand{\edm}{\end{displaymath}}

\def\a{\alpha}

\def\adot{{\dot\alpha}}

\newcommand{\im}{{\rm Im\,}}

\newcommand{\ie}{{\it i.e.\ }}
\newcommand{\eg}{{\it e.g.\ }}

\newcommand{\calR}{{\mathcal R}}

\DeclareMathAlphabet{\mathpzc}{OT1}{pzc}{m}{it}
\def\bea{\begin{eqnarray}}
\def\eea{\end{eqnarray}}
\def\beas{\begin{eqnarray*}}
\def\eeas{\end{eqnarray*}}
\def\sla{\raise.15ex\hbox{$/$}\kern-.57em}

\def\bea{\begin{eqnarray}}
\def\eea{\end{eqnarray}}

\def\sla{\raise.15ex\hbox{$/$}\kern-.57em}
\def\ie{{\it i.e.}~}
\def\eg{{\it e.g.}~}
\def\ap{{\alpha^\prime}}

\def\a{\alpha}

\def\vf{\varphi}

\def\cA{{\cal A}}
\def\cB{{\cal B}}

\def\cE{{\cal E}}
\def\cF{{\cal F}}
\def\cG{{\cal G}}
\def\cH{{\cal H}}
\def\cI{{\cal I}}

\def\cL{{\cal L}}

\def\cN{{\cal N}}

\def\cS{{\cal S}}
\def\cT{{\cal T}}
\def\cU{{\cal U}}
\def\cV{{\cal V}}

\def\cZ{{\cal Z}}

\begin{document}
\begin{titlepage}
\begin{flushright}
{ROM2F/2009/15}\\
NITheP-09-27
\end{flushright}
\vskip 2cm
\begin{center}
{\Large\bf Notes on un-oriented D-brane instantons\footnote{Based
on lectures delivered by M.Bianchi at the Fourth Young Researchers
Workshop of the European Superstring Theory Network in Cyprus,
September 2008.}}
\end{center}
\vskip 2cm
\begin{center}
{\large\bf Massimo Bianchi, \, Marine Samsonyan}~\\
{\sl Dipartimento di Fisica and Sezione I.N.F.N. \\ Universit\`a di Roma ``Tor Vergata''\\
Via della Ricerca Scientifica, 00133 Roma, Italy}\\

\end{center}
\vskip 3.0cm
\begin{center}
{\large \bf Abstract}
\end{center}

In the first lecture, we discuss basic aspects of worldsheet and
penta-brane instantons as well as (unoriented) D-brane instantons,
which is our main focus here, and threshold corrections to
BPS-saturated couplings. The second lecture is devoted to
non-perturbative superpotentials generated by `gauge' and `exotic'
instantons living on D3-branes at orientifold singularities. In
the third lecture we discuss the interplay between worldsheet and
D-string instantons on $T^4/Z_2$. We focus on a 4-fermi amplitude,
give Heterotic and perturbative Type I descriptions, and offer a
multi D-string instanton interpretation. We conclude with possible
interesting developments.



\vfill

\end{titlepage}

\section*{Introduction}

Aim of these lectures is introducing the interested reader to the
fascinating subject of non-perturbative effects generated by
unoriented D-brane instantons. We assume basic knowledge of
D-branes, Yang-Mills instantons and CFT techniques on the string
worldsheet. Due to lack of time and space we cannot but mention
recent developments such as wall crossing and localization.

In the first lecture, after a very short reminder of Yang-Mills
instantons and the ADHM construction, we discuss basic aspects of
worldsheet and D-brane instantons and their original applications
to threshold corrections.

The second lecture is devoted to non-perturbative superpotentials
generated by `gauge' and `exotic' instantons living on D3-branes
at orientifold singularities.

In the third lecture we discuss the interplay between worldsheet
and D-brane instantons on $T^4/Z_2$. We focus on a specific
4-hyperini amplitude, give Heterotic and perturbative Type I
descriptions, and offer a multi D-string instanton interpretation.
We conclude with proposing new perspectives and drawing lines for
future investigation.

Several good reviews are already available on the subject
\cite{BCKWreview}, that is also covered in textbooks
\cite{Kirbook}. We hope our short presentation could offer a
complementary view onto such an active research field.


\section{First Lecture: Instantons from Fields to Strings}

\subsection{Yang-Mills Instantons: a reminder}

Instantons (anti-instantons) are self-dual (anti-self-dual)
classical solutions of  the equations of motions of pure
Yang-Mills theory in Euclidean space-time. \bea F_{\mu\nu}=\pm
\tilde{F}_{\mu\nu} \label{dantd} \eea with
$\tilde{F}_{\mu\nu}=\frac{1}{2}\epsilon_{\mu\nu\rho\sigma}F_{\rho\sigma}
$. In quantum theory they can be thought of as gauge
configurations bridging quantum tunnelling among topologically
distinct vacua. It is remarkable that self-dual (anti-self-dual)
gauge fields automatically satisfy YM equations {\it in vacuo} as
a result of the Bianchi identities. These solutions are classified
by a topological charge: \bea K =\frac{g^2}{32
\pi^2}\int{d^4xF^a_{\mu\nu}\tilde{F}^a_{\mu\nu}}
\label{Pontryagin} \eea an integer, which computes how
many times an $SU(2)$ subgroup of the gauge group is wrapped by
the classical solution while its space-time location spans the
$S_3$-sphere at infinity. The action of a self-dual (or
anti-self-dual) instanton configuration turns out to be
$$S_I=\frac{8 \pi ^2}{g^2}|K|$$

\subsubsection{ADHM construction}

An elegant algebro-geometric construction of YM instantons was
elaborated by Atiyah, Drinfeld, Hitchin and Manin and goes under
the name of ADHM construction \cite{ADHM}.

For $SU(N)$ groups, the ADHM ansatz for a self-dual gauge field
with topological charge $K$, written as a traceless hermitean
$N\times N$ matrix, reads
$$(A_\mu)_{uv}(x)=g^{-1}\bar{U}_u^\lambda \partial_\mu U_{\lambda v} \quad ,$$
where $U_{\lambda u}(x)$ with $u=1,...,N $ and
$\lambda=1,...,N+2K$ are $(N+2K)\times N$ complex `matrices' whose
columns are the basis ortho-normal vectors for the $N$ dimensional
null-space of a complex $2K\times (N+2K)$ `matrix'
$\bar{\Delta}(x)$, \ie satisfy
$$\bar{\Delta}_i^{\dot{\alpha}\lambda}U_{\lambda
u}=0=\bar{U}_u^\lambda \Delta_{\lambda i \dot{\alpha}}$$ for
$i=1,...,K$, $\alpha, \dot{\alpha}=1,2$. Remarkably,
$\Delta_{\lambda i\dot{\alpha}}(x)$ turns out to be at most linear
in $x$. In quaternionic notation\footnote{Any real 4-vector
$V_\mu$ can be written as a `real' quaternion $V_{\a\dot\a} =
V_\mu \sigma^\mu_{\a\dot\a}$ with $\sigma^\mu = \{1,-i\sigma^a\}$.}
for $x$,
$$\Delta_{\lambda i\dot{\alpha}}(x)=
a_{\lambda i\dot{\alpha}}+b^{\alpha}_{\lambda
i}x_{\alpha\dot{\alpha}}
 \quad , \quad \bar{\Delta}_i^{\dot{\alpha}\lambda}
(x)=\bar{a}_i^{\dot{\alpha}\lambda}+
\bar{x}^{\dot{\alpha}\alpha}\bar{b}_{i\alpha}^{\lambda}\equiv(\Delta_{\lambda
i\dot{\alpha}})^* .$$ The complex constant `matrices' $a$ and $b$
form a redundant set of collective coordinates that include the
moduli space ${\cal{M}}_{K}$. Decomposing the index
$\lambda$ as $\lambda=u+i\alpha$, with no loss of
generality, one can choose a simple canonical form for $b$
$$b^\beta_{\lambda j}=b^\beta_{(u+i\alpha)j}=\left(\begin{array}{c}
0\\
\delta_{\alpha}^{\beta} \delta_{ij}
\end{array}
\right), \quad \bar{b}_{\beta j}^\lambda=\bar{b}^{(u+i\alpha)}
_{\beta j}=\left(\begin{array}{c} 0 \quad \delta _\alpha ^\beta
\delta _{ij}
\end{array}
\right)$$ One  can also split $a$ in a similar way as:
$$a_{\lambda j \dot{\alpha}}=a_{(u+i\alpha)j\dot{\alpha}}
=\left(\begin{array}{cc} w_{uj \dot{\alpha}}\\
(X_{\alpha \dot{\alpha}})_{ij}
\end{array}
\right), \quad \bar{a}_j^{\dot{\alpha}\lambda}=
\bar{a}^{\dot{\alpha}(u+i\alpha)}_j= \left(\begin{array}{c}
\bar{w}_{j}^{\dot{\alpha}u} \quad (\bar{X}^{\dot{\alpha}\alpha
})_{ji}
\end{array}
\right)$$ In order to ensure self-duality of the connection, the
`ADHM data' $\{ w, \bar{w}, X, \bar{X}\}$ with
$X_\mu^{\dagger}=X_\mu$ must satisfy algebraic constraints, known
as the ADHM equations, that can be written in the form
$$w_{ui \dot{\alpha}} (\sigma^a)^{\dot\alpha}{}_{\dot\beta}
\bar{w}_{j}^{\dot{\beta}u} + \eta^a_{\mu\nu} [X^\mu,X^\nu]_{ij} =
0$$ for later comparison with the D-brane construction. Note the $U(K)$ invariance of the above $3K\times K$ equations.
For a recent review of supersymmetric instanton calculus see \cite{MKR}.

The ADHM construction for unitary groups can be generalized to
orthogonal and symplectic groups. It is quite remarkable how the
rather abstract ADHM construction can be made very intuitive using
D-branes and $\Omega$-planes \cite{Douglas'95} as we will see
later on.

\subsection{Instantons in String Theory}

\subsubsection{Worldsheet instantons}

World-sheet instantons in Heterotic and Type II theories
correspond to Euclidean fundamental string world-sheets wrapping
topologically non-trivial internal cycles of the compactification
space and produce effects that scale as $e^{-R^2/\ap}$
\cite{Dine}. Depending on the number of supersymmetries (thus on
the number of fermionic zero modes), they can correct the
two-derivative effective action or they can contribute to
threshold corrections to higher derivative (BPS saturated)
couplings \cite{DualitInst}. For Type II compactifications on CY
three-folds, preserving $\cN=2$ supersymmetry in $D=4$,
holomorphic worldsheet instantons ($\bar\partial X = 0$) correct
the special K\"ahler geometry of vector multiplets (Type IIA) or
the dual quaternionic geometry of hypermultiplets (Type IIB). For
heterotic compactifications with standard embedding of the
holonomy group in the gauge group, complex structure deformations
are governed by the same special K\"ahler geometry as in Type IIB
on the same CY three-fold, that is not corrected by worldsheet
instantons. Complexified K\"ahler deformations are governed by the
same special K\"ahler geometry as in Type IIA on the same CY
three-fold, that is corrected by worldsheet instantons, or
equivalently, as a result of mirror symmetry, by the same special
K\"ahler geometry as in Type IIB on the mirror CY three-fold that
is tree level exact. For standard embedding, the K\"ahler metrics
of charged supermultiplets in the $27$ and $27^*$ representations
of the surviving/visible $E_6$ are simply determined by the ones
of the neutral moduli of the same kind by a rescaling \cite{25}.
For non standard embeddings the situation is much subtler.

\subsubsection{Brane instantons} Euclidean NS5-branes (EN5-branes)
wrapping the 6-dimensional compactification manifold produce
non-perturbative effects in $e^{-c/g_s^2}$ (reflecting the
NS5-brane tension) that qualitatively correspond to `standard'
gauge and gravitational instantons. Euclidean Dp-brane wrapping
$(p+1)$-cycles produce instanton effects that scale as
$e^{-c_p/g_s}$  (reflecting the EDp-brane tension)\cite{Becker}.
In Type IIB on CY three-fold, ED(-1), ED1-, ED3- and ED5-brane
instantons, obtained by wrapping holomorphic submanifolds, correct
dual quaternionic geometry in combination with world-sheet (EF1-)
and EN5-instantons. In Type IIA on CY three-folds, ED2-instantons
(D-`membrane' instantons) wrapping special Lagrangian
submanifolds, correct the dual quaternionic geometry, in
combination with EN5-instantons. In both cases, the dilaton
belongs to the universal hypermultiplet.

\subsubsection{Unoriented D-brane instantons} In Type I, the
presence of $\Omega$9-planes  severely restricts the possible
homologically non trivial instanton configurations. Only ED1- and
ED5-branes are homologically stable. Other (Euclidean) branes may
be associated to instanton with torsion (K-theory) charges. For
other un-oriented strings the situation is similar and can be
deduced by means of T-duality: e.g. for intersecting D6-branes one
has two different kinds of ED2-branes (ED0- and/or ED4-brane
instanton require $b_{1,5}\neq 0$), for intersecting D3- and D7-
branes one has ED(-1) and ED3-branes. There are two classes of
unoriented D-brane instantons depending on the stack of branes
under consideration.
\begin{itemize}
\item{ `Gauge' instantons correspond to EDp-branes wrapping
the same cycle $\cal C$ as a stack of background D(p+4)-branes. The
prototype is the D3, D(-1) system \cite{Billo} that has 4 N-D
directions. The EDp-branes behave as instantons inside D(p+4)-branes
$$F=\tilde{F}$$
and produce effects whose strength, given by
$$e^{-W_{p+1}(C)/g_s \ell_s^{p+1}}=e^{-1/g_{YM}^2},$$ }
is precisely the one expected from `gauge' instantons in the
effective field-theory.

 \item{ `Exotic' instantons arise from EDp'-branes wrap a cycle
 ${\cal C}'$ which is not wrapped by any stack of background D(p+4)-branes. The prototype is
the D9, ED1 system with 8 N-D directions and only a chiral fermion
at the intersection. In this case
$$F\neq \tilde{F}$$
and the strength is given by
$$e^{-W_{p'+1}(C')/g_s \ell_s^{p'+1}} \neq e^{-1/g_{YM}^2}$$
`Exotic' instantons may eventually enjoy a field theory
description in terms of octonionic instantons or hyper-instantons
with $F\wedge F = *_8 F\wedge F$.}
\end{itemize}

\subsection{Original Applications and Various Comments}

Let us now list possible effects generated by (un)oriented D-brane
instantons in diverse string compactifications.

\begin{itemize}

\item{In $\cN =8$ theories (\eg toroidal compactifications of
oriented Type II A/B) D-brane instantons produce threshold corrections to $R^4$
terms and other 1/2 BPS (higher derivative) terms}.

\item{In $\cN =4$ theories (\eg toroidal compactifications of Type I
/ Heterotic) D-brane instantons produce threshold corrections to $F^4$ terms
and other 1/2 BPS (higher derivative) terms}.

\item{In $\cN =2$ theories (\eg toroidal orbifolds with
$\Gamma\subset SU(2)$) D-brane instantons produce threshold corrections to
$F^2$ terms and other 1/2 BPS terms}.

\item{In $\cN =1$ theories (\eg toroidal orbifolds with
$\Gamma\subset SU(3)$) D-brane instantons produce threshold corrections and
superpotential terms}.

\end{itemize}

\subsubsection{Thresholds in toroidal compactifications}

We have not much to add to the vast literature on threshold
corrections to $R^4$ terms in $\cN =8$ theories which are induced
by oriented D-brane as well as world-sheet instantons\footnote{In
$D=4$ and lower Euclidean NS5-branes can also contribute.}. We
would only like to argue that in unoriented Type I strings and
alike these corrections should survive as functions of the
unprojected closed string moduli despite some of the corresponding
D-brane or worldsheet instantons be not BPS. These and lower
derivative ($R^2$) couplings may receive further perturbative
corrections from surfaces with boundaries and crosscaps. Viz: ${\cal L}_{II}\approx {\calR}^4f_{II}(\phi ,
\chi)\rightarrow {\cal L}_I\approx {\calR}^4[f_{II}(\phi , \chi=0)+f_I(\phi)]$.

The original application of unoriented D-brane instanton was in
the context of threshold corrections to $F^4$ terms in
toroidal compactifications of Type I strings \cite{BKB}. These are closely related to the threshold corrections to $F^4$
terms for heterotic strings on $T^d$. For later use, let us briefly summarize the structure of the latter.
After
\begin{itemize}
\item{Computing the one-loop correlation function of 4 gauge boson vertex operators $V_{(0)}=A_\mu ^a(\partial
X^\mu+ip\psi\psi^\mu)\tilde{J}_ae^{ipx}$}
\item{Taking the limit of zero momentum in the exponential factors \ie neglecting
the factor $\Pi(z_i,p_i) = \prod_{i,j} exp[-\ap p_i\cdot p_j \cG(z_{ij})] \rightarrow 1$}
\end{itemize}
or, equivalently,
\begin{itemize}
\item{computing the character-valued partition function in a
constant field-strength background $\nu$} \item{taking the fourth
derivative wrt $\nu$}
\end{itemize}
one arrives at the integral over the one-loop moduli space that
receives contribution only from BPS states and schematically reads
$$\cI_{d} [\Phi]= \cV_d \int_{\cF} {d^2\tau \over \tau_2^2} \sum_M
e^{2\pi i \cT(M)} e^{-{\pi Im\cT(M)\over \tau_2 Im\cU(M)}|\tau -
\cU(M)|^2} \Phi (\tau)$$
where $M=(\vec{n}, \vec{m})$ represent the embedding of the world-sheet torus in the target $T^d$, $\Phi(\tau)$ is some
modular form. The induced Kahler $\cT(M)$ and complex $\cU(M)$ structures are given by
$$\cT(M) = \cB_{12} + i \sqrt{\det\cG} \quad, \quad \cU(M) = {1\over \cG_{11}} (\cG_{12} + i \sqrt{\det\cG}) $$
with $ \cG = M^t G M $, $\cB = M^t B M $ induced metric and
$B$-field \cite{Kiritsis}. The integral can be decomposed into
three terms $\cI_{d}[\Phi]= \cI_{d}^{triv}[\Phi] +
\cI_{d}^{deg}[\Phi] + \cI_{d}^{ndeg}[\Phi]$.  The three different
orbits are classified as follows: the orbit of $M=0$ (trivial
orbit), degenerate orbits with $\det(M^{i,j})=0$ and
non-degenerate orbits with some $\det(M^{i,j})\neq 0$. Let us
consider the various contributions.
\begin{itemize}
\item {\bf Trivial orbit}: $M=0$,
$$\cI_{d,d}^{triv}[\Phi] =
\int_{\cF}{d^2\tau \over \tau_2^2} \Phi(\tau) \rightarrow
\cI_{d,d}^{triv}[1] = {\pi^2 \over 3} \cV_d $$

\item {\bf Degenerate orbits}: $M\neq 0$, $\det(M^{i,j}) = n^i m^j
- n^j m^i =0$ $\forall i,j$. One can choose $\vec{n}=0$
representative and unfold $\cF$ to the strip $\cS = \{
|\tau_1|<1/2, \tau_2>0\}$, then

$$ \cI_{d,d}^{deg}[\Phi] = \cV_d \int_{\cS}{d^2\tau \over
\tau_2^2}\sum_{\vec{m} \neq \vec{0}} e^{-{\pi\over \tau_2}
\vec{m}^t G \vec{m}} \Phi(\tau) \rightarrow \cI_{d}^{deg}[1]= \cV_d
\cE^{SL(d)}_{\bf d}(G).$$

\item {\bf Non degenerate orbits}: at least one $\det(M_{ij}) =
n^i m^j - n^j m^i \neq 0$. The representative for these orbits may
be chosen to be $\vec{n}^\alpha=0$ for $\alpha = 1, .., k$,
$m^\alpha \neq 0$, $n^{\bar\alpha}> m^{\bar\alpha}\ge 0$ and
enlarging the region of integration $\cF$ to the full upper half
plane $\cH^+$ one finds:
$$ \cI_{d,d}^{ndeg}[\Phi] = \cV_d
\int_{\cH^+} {d^2\tau \over \tau_2^2}
\sum_{(n^{\bar\a},0;m^{\bar\a},m^{\a})} e^{2\pi i \cT(M)} e^{-{\pi
Im\cT(M)\over \tau_2 Im\cU(M)}|\tau - \cU(M)|^2} \Phi(\tau)$$
$$ \rightarrow \cI_{d,d}^{deg}[1]= \cV_d \cE^{SO(d,d)}_{{\bf
V}, s=1}(G,B) \quad { \rm {(generalized \, Eisenstein \, series)}}. $$

\end{itemize}

Thanks to Type I / Heterotic duality, heterotic worldsheet
instantons are mapped into ED-string instantons. Since $F^4$ terms
are 1/2 BPS saturated, matching the spectrum of excitations,
including their charges, was believed to be sufficient to match
the threshold corrections even in the presence of (non)commuting
Wilson lines \cite{BKB,MBEGJFMKN} or after T-duality
\cite{Gutperetal, KFSS}. More recently, thanks to powerful localization
techniques, a perfect match between threshold corrections in
Heterotic and Type I'' (with D7-branes) has been found on $T^2$
for the specific choice of commuting Wilson lines breaking
$SO(32)$ to $SO(8)^4$ \cite{Lerdaetal}. The somewhat
unsatisfactory results of \cite{FMP} for different breaking
patterns with orthogonal or symplectic groups can be either
interpreted as a failure of localization or as the need to include
higher order terms. Notice that only for $SO(8)$, `exotic'
string instantons should admit a field theory interpretation in
terms of `octonionic' instantons. It would be nice to further
explore this issue in this or closely related context of $\cN=1,
2$ theories in D=4 where heterotic worldsheet instantons
correcting the gauge kinetic function should be dual to ED-string
(or other ED-brane) instantons \cite{ABDFPT}. A short review of the strategy to
compute similar threshold corrections will be presented later
on when we discuss Heterotic / Type I duality on $T^4/Z_2$.

\subsubsection{Phenomenological considerations}

Despite some success in embedding (MS)SM in vacuum configurations
with open and unoriented strings, there are few hampering
properties at the perturbative level:

\begin{itemize}
\item{Forbidden Yukawas in $U(5)$ (susy) GUT's
$$H^d_{{\bf 5}^*_{-1}}F^c_{{\bf 5}^*_{-1}}A_{{\bf 10}_{+2}} \quad
OK \quad {\bf but} \quad H^u_{{\bf 5}_{+1}} A_{{\bf 10}_{+2}}
A_{{\bf 10}_{+2}}  \quad KO $$ forbidden by (global, anomalous)
$U(1)$ invariance, though compatible with $SU(5)$ (yet no way
$\eps^{abcde}$ from Chan-Paton)}

\item{R-handed (s)neutrino masses $W_M=M_R NN$ forbidden by \eg
$U(1)_{B-L}$ in Pati-Salam like models $SO(6)\times SO(4)
\rightarrow SU(3)\times SU(2)_L\times SU(2)_R \times U(1)_{B-L}$}

\item{$\mu$-term in MSSM $W_{\mu}=\mu H_1 H_2$ typically forbidden
by extra (anomalous) $U(1)$'s}
\end{itemize}

All the above couplings can be generated by `stringy' instantons
after integrating over the `non-dynamical' moduli living on the
world-volume of the EDp'-branes under consideration. These effects
scale as $e^{-T_{EDp'} V_{EDp'}}$ and are non-perturbative in
$g_s$, since $T_{EDp'}\approx 1/g_s (\ap)^{p+1/2}$. Yet {\it a
priori} they depend on different moduli (through the dependence of
$V_{EDp'}$ on variuos $Z$'s) from the ones appearing in the gauge
kinetic function(s) of background Dp'-brane, so they cannot in
general be identified with the standard `gauge' instantons.
Relying on the $g_s$ power counting introduced in \cite{Billo},
the relevant are disks with insertions of the non-dynamical vertex
operators $V_\Theta$ (connecting EDp'-EDp') and $V_\lambda$
(connecting EDp-Dp') with or without insertions of dynamical
vertex operators $V_A$ etc, which correspond to the massless
excitations of the vacuum configuration of
(intersecting/magnetized) unoriented Dp-branes \cite{Blum}. Disks
without dynamical insertions yield the `instanton action', with
one dynamical vertex they produce classical profiles for $A$ etc.
Disks with more insertions contribute to higher-order corrections.
One loop diagrams with no insertions produce running couplings and
subtle numerical prefactor that can cancel a given type of
non-perturbative F-terms \cite{Silverstein, BWnieist}.

\subsubsection{Anomalous $U(1)$'s and gauged PQ symmetries}

In general, a `naked' chiral field $Z$  whose pseudoscalar axionic
components $\zeta=\im Z$ shift under some local anomalous $U(1)$
cannot appear in a (super)potential term if not dressed with other
chiral fields charged under $U(1)$. $U(1)$ invariance puts tight
constraints on the form of the possible superpotential terms.
Since the axionic shift is gauged it must be a symmetry of the
kinetic term. This is only possible when no non-perturbative
(world-sheet or D-brane instanton) corrections spoil the tree
level (in fact perturbative) PQ symmetry. This means that the
gauging procedure corresponds to turning on fluxes such that the
potential instanton corrections in $Z$ are in fact disallowed. In
practice, this means the corresponding wrapped brane is either
anomalous (\`a la Freed-Witten) \cite{FW} or destabilized due to
the flux \cite{KPT}.

Moreover, background fluxes (for both open and closed strings) can
lift fermionic zero-modes. Various `perturbative' studies have
been carried out \cite{instflux}.


\section{Second Lecture: Unoriented D-brane Instantons}

\subsection{ADHM from branes within branes}

As already mentioned, the ADHM construction has a rather intuitive
description in open string theory, whereby the gauge theory is
realized on a stack of Dp-branes. D(p-4)-branes which are
localized within the previous stack of branes behave as instantons
\cite{Douglas'95}.

Indeed, the WZ couplings on the Dp-brane worldvolume schematically
reads \be S_{WZ} = \int C_{p+1} + \int C_{p-1} \wedge Tr(F) + \int
C_{p-3} \wedge Tr(F\wedge F) + ... \ee

In particular a localized source of $C_{p-3}$ within a Dp-brane
behaves like an instanton density $Tr(F\wedge F)$. Moreover, the
ADHM data are nothing but the massless modes of open strings
connecting the D(p-4)-branes with one another or with the
background Dp-branes.

Let us take $p=3$ for definiteness. The low-energy effective
theory on the world-volume of $N$ parallel D3-branes is $N=4$
supersymmetric Yang-Mills theory with gauge group $U(N)$.
Instanton moduli are described by the massless modes of open
strings with at least one end on the D(-1)-brane stack.  In this
system of D3 and D(-1) branes there are three sectors of the open
string spectrum to be considered. $U(N)$ gauge fields and their
superpartners are provided by the strings that start and end on
D3-branes. Strings stretching between two D(-1)-branes give rise
to $U(K)$ non-dynamical gauge fields and their superpartners.
These `fields' represent part of the (super) ADHM data. The
remaining (super) ADHM data are provided by the strings with one
end on the D3-branes and the other one on the D(-1)-branes and
{\it vice-versa}.

In the presence of D3-branes, (Euclidean) Lorentz symmetry is
broken $SO(10)\rightarrow SO(4)\times SO(6)$ and it is convenient
to split ten ``gauge bosons'', $A_M$, into four gauge bosons
$a_\mu$, and six real ``scalars", $\chi_i$. Similarly the $d=10$
gauginos produce four non-dynamical Weyl ``gauginos'',
$\Theta_\alpha^A$ as well as their antiparticles
$\bar{\Theta}_A^{\dot{\alpha}}$.

Introducing, for later convenience, three auxiliary fields $D^c$,
the D(-1)-D(-1) $U(K)$ `geometric' supermoduli are given by
$$a_\mu,\quad \chi_i;\quad
\Theta_\alpha^A,\quad \bar{\Theta}_A^{\dot{\alpha}};\quad D^c$$
while the $4KN$ D(-1)-D(3) `gauge' supermoduli are
$$w_{\dot{\alpha}i}^u,\quad \bar{w}_{\dot{\alpha}u}^i;\quad
\nu_{i}^{Au},\quad \bar{\nu}_{u}^{Ai}$$ with $\mu=1,...,4,\quad
\alpha,\dot{\alpha}=1,2$ vector and spinor indices of $SO(4)$,
$i=1,...,6,\quad A=1,...,4$ are vector and spinor indices of
$SO(6)$ respectively and $c=1,2,3$. The matrices $a_\mu,\, \chi_i$
describe the position of the instanton along the longitudinal and
transverse directions to the D3-brane respectively.
$\Theta_\alpha^A$ and $\bar{\Theta}_A^{\dot{\alpha}}$ are their
superpartners. $w_{\dot{\alpha}}$, $\bar{w}_{\dot{\alpha}}$
represent the D3-D(-1) open string in the $NS$ sector, accounting
for instanton sizes and orientations, and $\nu^A$, $\bar{\nu}^A$
are their fermionic superpartners.

\subsection{The D3-D(-1) action}

By computing scattering amplitudes on the disk, one can determine
the complete action that governs the dynamics of the light modes
(or moduli) of the system of D(-1) branes in the presence of
D3-branes. It schematically reads \cite{Dorey}
$$S_{K,N}={Tr}_{k}\left[{1\over g_0^{2}}S_{G}+S_{K}+S_{D}\right]$$
with
$$S_{G}=-[\chi_i,\chi_j]^2+ i\bar{\Theta}_{\dot{\alpha}
A}[\chi_{AB}^\dagger,\bar{\Theta}^{\dot{\alpha}}_B]-D^{c}D^{c}$$
$$S_{K}=-[\chi_i,a_{\mu}]^2+\chi^i\bar{w}^{\adot}
w_{\adot}\chi_i-i\Theta^{\alpha A}[\chi_{AB},
\Theta^{B}_{\alpha}]+2i\chi_{AB}\bar{\nu}^{A}\nu^{B}$$
$$S_{D}=i\left(-[a_{\alpha\dot{\alpha}},\Theta^{\alpha A}]
+\bar{\nu}^A w_{\dot{\alpha}} +\bar{w}_{\dot{\alpha}}\nu^A \right)
\bar{\Theta}^{\dot{\alpha}}_A+D^{c}\left(\bar{w}\sigma^c w-i
\bar{\eta}_{\mu\nu}^c [a^\mu,a^\nu] \right)$$ where
$\chi_{AB}\equiv{1\over 2}\Sigma^i_{AB}\chi_i$ and
$\Sigma^i_{AB}=(\eta^c_{AB},i\bar{\eta}^c_{AB})$ are given in
terms of t'Hooft symbols and
$g_0^2=4\pi(4\pi^2\alpha^{\prime})^{-2}g_s$. Note that the action
$S_{K,N}$ arises from the dimensional reduction of the D5-D9
action in six dimensions down to zero dimension. If there are
v.e.v. for the six $U(N)$-adjoint scalars $\varphi_a$ belonging to
the D3-D3 open string sector one has to add the term
$$S_\varphi=tr_{k}\left[\bar{w}^{\dot{\alpha}}(\varphi^i \varphi_i+2
\chi^i\varphi_i)w_{\dot{\alpha}}+2i\bar{\nu}^A \varphi_{AB}\nu^B
\right]$$ to the action $S_{K,N}$. In the limit $g_0\sim
(\alpha^{\prime})^{-1}\rightarrow \infty$ ($g_0$ fixed) gravity
decouples from the gauge theory and there are no contributions
coming from $S_G$. $\bar{\Theta}_{\dot{\alpha}A}$ and $D^c$ fields
become Lagrange multipliers for the super ADHM constraints:
$$D^a:\quad [a_\mu,a_\nu]\eta_a^{\mu\nu}+w\sigma_a\bar{w}=0 \qquad {\rm ADHM \quad Eqs}$$
$$\bar{\Theta}_A^{\dot{\alpha}}:\quad
[a_\mu,\Theta^A]\sigma^{\mu}+w\bar{\nu}^A+\nu^A\bar{w}=0\qquad
{\rm super \quad ADHM \quad Eqs}$$

In this limit the multi-instanton `partition function' becomes
$$\cZ_{k,N}=\int_{\mathfrak{M}}e^{-S_{k,N}-S_\vf}=\frac{1}{{\rm
Vol}\,U(k)}\, \int_{\mathfrak{M}} d\chi\, dD\,da\,
d\theta\,d\bar{\theta} dw\, d\nu\, e^{-S_{k,N}-S_\vf}.$$

\subsection{Vertex operators}
Classical actions, (super)instanton profiles and non-perturbative
contributions to scattering amplitudes can be derived by computing
disk amplitudes with insertions of vertex operators for
non-dynamical moduli $V_a$, $V_\chi$, $V_w$, $V_{w^\dagger}$ (ADHM
data) and their superpartners \cite{Billo}.

\subsubsection{Vertex operators for `gauge' instantons}

Let us first start considering the vertex operator for a non
dynamical gauge boson $a_\mu$ along the four D-D space-time
directions. The vertex operator reads
$$V_a=a_\mu e^{-\varphi}\psi^\mu T_{K\times K}$$
where $\vf$ arises from the bosonization of the $\beta, \gamma$
worldsheet super-ghosts, $\psi$ are the worldsheet fermions and
$T_{K\times K}$ are $U(K)$ Chan-Paton matrices. For the non
dynamical transverse scalars $\chi_i$ along the six internal D-D
directions, the vertex operator reads
$$V_\chi=\chi_i e^{-\varphi}\psi^i T_{K\times K}$$
Similarly
$$V_\Lambda=\Theta^a(p) S_a e^{-\varphi / 2} T_{K \times K}$$
with $a=,1...,16$, produces four non-dynamical Weyl ``gauginos'',
$\Theta^A_\alpha$, and their antiparticles,
$\bar{\Theta}_A^\adot$.

Bosonic vertex operators for low-lying D(p-4)-Dp strings, with
multiplicity $K\times N$ and their conjugates are given by
$$V_w=\sqrt{g_s\over v_{p-3}}w_\alpha e^{-\varphi}\prod_\mu\sigma_{\mu}S^\alpha T_{K\times N}$$
with $S^\alpha$ an $SO(4)$ spin field of worldsheet scaling
dimension $1/4$. $\sigma_\mu$ are $Z_2$ bosonic twist fields along
the 4 relatively transverse N-D directions. $\Pi_\mu \sigma_\mu$
has total dimension $1/4=4/16$. $T_{K,N}$ denote the $K\times N$
Chan-Paton `matrices'. The super-partners of $w_\alpha$ are
represented by vertex operators of the form
$$V_\nu=\sqrt{g_s\over v_{p-3}}\nu_A e^{-\varphi / 2}\prod_\mu\sigma_{\mu}S^A T_{K\times N}$$
where $S^A$ is an $SO(6)$ spin field of dimension 3/8. Note that
the overall normalization $\sqrt{g_s/ v_{p-3}}$ is crucial for the
correct field theory limit $(\alpha^\prime\rightarrow 0)$.

\subsubsection{Vertex operators for 'stringy' instantons}

Let us now consider `stringy' instantons. The prototype is the D9,
D1 system which has 8 N-D directions. The multi-(instanton)
configuration of this system was first analyzed in \cite{BKB}. The
lowest lying modes of an open string stretched between N D9 and K
D1 branes are massless fermions with a given chirality (say Right)
along the two common N-N directions. For Type I strings there are
32 such chiral fermions ($\lambda^A$) that precisely reproduce the
gauge degrees of freedom of the `dual' heterotic string
\cite{WitPol}. In addition, in the $\cN = (8,0)$ theory on the D1
world-sheet with $SO(8)$ R-symmetry group, there are 8 transverse
bosons $X^I$ in the $8_v$ and as many Green-Schwarz type fermions
$S^a$ of opposite chirality (say Left) in the $8_s$ of $SO(8)$.
The 32 massless right-moving $\lambda^A$ are inert under the
left-moving susy $Q_{\dot{a}}$ in the $8_c$.

After compactification to $D=4$ on a manifold with non-trivial
holonomy some of the global supersymmetries are broken and the
corresponding D1 world-sheet theory changes accordingly. In
particular $SO(8)$ breaks to some subgroup.

\subsection{D-branes at Orbifolds}

A particularly promising class of configurations with nice phenomenological perspectives that also allow explicit
non-perturbative computations are unoriented D-branes at singularities.
Let us consider a stack of D3-branes at the orbifold singularity
$T^d/\Gamma \approx R^d/\Gamma$ (locally), and let us take $\Gamma
=Z_n$ for simplicity. At the singularity $N$ D3-branes group into
stacks of $N_i$ `fractional' branes, that cannot move away from
the singularity, with $i=0,1,2,...$ labelling the conjugacy
classes of $Z_n$. The gauge group $U(N)$ decomposes as $\Pi_i
U(N_i)$. \bea (Z_1, Z_2, Z_3)\approx(\omega^{k_1}Z_1,\omega^{k_2}Z_2,\omega^{k_3}Z_3)
\label{proj} \eea for simplicity we take $k_1+k_2+k_3=0~(mod~n)$ that generically preserves $\cN=1$ supersymmetry.

The action on Chan-Paton factors is given by
$$\rho(Z_n)=\rho_0(1_{N_0}, \omega^{1} 1_{N_1},\omega^{2}
1_{N_2}, ..., \omega^{n-1} 1_{N_{n-1}})$$ For $\ap \approx 0$,
keeping only invariant components under (\ref{proj}), the
resulting theory turns out to be an $\cN=1$ quiver gauge theory,
in which vector multiplets $V$ are in the $N_i \bar{N}_i$
representation while chiral multiplets $\Phi_i$ are in the $N_j
\bar{N}_l$ representation with $k_i + j - l = 0~(mod~n)$
\cite{U(4)}.

Twisted RR tadpole cancellation in sectors with non vanishing
Witten index can be written as $tr\rho(Z_n)=0$ that ensures the
cancellation of chiral non-abelian anomalies \cite{Anom&Tad}.

\subsubsection{Unoriented projection}

Possible unoriented projections depend on the parity of $n$ and
the charge of the $\Omega$-plane. For $n$ odd there is only one
possibility
$$ N_0=\bar{N}_0\quad,\quad N_{i}=\bar{N}_{n-i} $$
For $n$ even there are two possibilities
$$ N_0=\bar{N}_0\quad , \quad N_{i}=\bar{N}_{n-i}\quad , \quad N_{n/2} = \bar{N}_{n/2}$$
$$ N_0=\bar{N}_{n/2} \quad , \quad N_{i}=\bar{N}_{n/2-i} $$
One should also impose the twisted RR tadpole cancellation condition (non vanishing Witten index)
$tr\rho(Z_n)=\pm q_n^\Omega$ which from the field theory point of view is just the chiral anomaly
cancellation \cite{AIMU}.

Let us focus on the very rich and instructive case of $T^6/Z_3
\approx R^6/ Z_3$.

\subsection{Unoriented $R^6/Z_3$ projection}

In the remaining part of this Section, for illustrative purposes,
we will discuss unoriented D-brane instantons on a stack of
D3-branes located at an unoriented $R^6/Z_3$ orbifold singularity.

Since $n=3$ is odd, there is only one possible embedding in the
Chan-Paton group up to the charge of the $\Omega 3^{\pm}$ planes.
Introduction of $\Omega 3^-$-plane combined with local R-R tadpole
cancellation leads to a theory with gauge group $G=SO(N_0)\times
U(N_0+4) \times H_{reg}$, where $H_{reg}$ accounts for the
Chan-Paton group of the `regular' branes that can move into the
bulk. We will henceforth assume that regular branes are far from
the singularity and essentially decoupled from the local quiver
theory. For $N_0 = 0$, we have $U(4)$ gauge group with 3 chirals
in ${\bf 6}_{-2}$. In the presence of $\Omega 3^+$-plane we get a
theory with $G=Sp(2N_0)\times U(2N_0-4) \times H_{reg}$ gauge
group, \eg for $2N_0=6$, we have $Sp(6)\times U(2)$ gauge group
with 3 chirals in $({\bf 6}, {\bf 2}_{+1})+({\bf 1}, {\bf
3}_{-2})$.

In both cases the anomalous $U(1)$ mixes with the twisted RR axion
$\zeta$ in a closed string chiral (linear) multiplet $Z$ (gauging
of axionic shift)
$$ \delta A=d\alpha \quad , \quad \zeta =-M_A \alpha $$
$$ \cL_{ax}=(d\zeta-M_AA)^2+{1\over f_\zeta}\zeta \: F\wedge F $$
Anomaly cancellation $\delta_\alpha [\cL_{ax}+\cL_{1-loop}]=0
\leftrightarrow M_A/f_\zeta=t_3=Tr_f Q^3$
\footnote{More complicated cases with several (non)anomalous $U(1)$'s, generalized Chern-Simons
terms $\cL_{GCS}=E_{[ij]k}A^i\wedge A^j \wedge F^k$
are needed in the low-energy effective theory with non-trivial
phenomenological consequences \cite{Lionettoetal}.}.

\subsubsection{Field Theory analysis}

As already mentioned `gauge' instantons are expected to generate VY-ADS-like superpotentials.
Neglecting  U(1)'s for the time being, the two choices of
$\Omega$-planes and, thus, of gauge group lead to superpotentials
of the form
$$SU(4) \quad : \quad  W={\Lambda^9 \over \det_{I,J}(\eps_{abcd}A^{ab}_I A^{cd}_J)}$$
$$Sp(6)\times SU(2) \quad : \quad  W={\Lambda^9 \over \det_{6\times 6}(\Phi^i_{aI})}$$
In string theory, $\Lambda^\beta=M^\beta_s e^{-{S\over f_a}-{Z\over
f_\zeta}}$ ($\beta=9$ here), shift of $Z$ compensates the $U(1)$
charge of the denominator!
The {\it ``thumb rule''} is that in each case there are two
exact/unlifted fermionic zero-modes $n(\lambda_0) - n(\psi_0) = 2
$. The rest is lifted by Yukawa interaction $Y_g=g \phi^\dagger
\psi\lambda$.
We now pass to describe the explicit computations with unoriented
D-instantons
$$ U(4)_{D3}\rightarrow U(K)_{D(-1)}
 \quad , \quad Sp(6)_{D3} \rightarrow O(K)_{D(-1)}$$
In both cases, there are two exact un-lifted fermionic zero-modes
for $K=1$.

\subsubsection{Non-perturbative superpotential for $Sp(6)\times
U(2)$}

After the projection, in the D(-1)-D(-1) sector one has geometric
supermoduli: $a_\mu$ (instanton position) and $\Theta^0_\alpha$
(Grassman coordinate), which yield the $\cN =1$ superspace
measure. There is no room for $D^c$ and
$\bar\Theta_0^{\dot\alpha}$ in the present case, since the
relevant instanton is an $O(1)$ instanton in the $Sp(6)$ group and
as such there are no super ADHM constraints.

In the D(-1)-D3 sector the gauge super-moduli are
$w^u_{\dot\alpha} \: , \nu^{0u}\: , \nu^{Ia}$ with $u=1,...6$
$Sp(6)$, $a=1,2$ $U(2)$, and $I=1,2,3$ $SU(3)$ `pseudo' flavor
indices\footnote{In string theory, $SU(3)$ is an accidental
symmetry of the two-derivative effective action}. Both
$\bar{\Theta}_{0 \adot}$ and $D^c$ are projected out. Taking into
account the interactions with the D3-D3 excitations
$\Phi^{Iua}=\phi^{Iua}+... $, the instanton action can be reduced
to the form
$$S_{D(-1)-D(3)}=w^u_{\dot\alpha}\bar\phi_{Iua}\phi^{Iva}w_{v\dot\alpha}+
\nu^{0u} \nu^{Ia} \bar\phi_{u Ia} $$
Integrations over gauge super-moduli are gaussian and the final
result can be written as $$ \int d^6w d^3\nu^0 d^3\nu^I
e^{S_{D(-1)-D(3)}}= {\det(\bar\phi_{u,Ia})\over
\det(\bar\phi_{u,Ia}\phi^{v,Ia})}={1 \over \det(\phi^{v,Ia})} $$
Including D(-1)-D(-1) action and one-loop contribution, up to a
non vanishing numerical constant, we get $$ \int d^4a d^2\Theta
{\mu^9 e^{2\pi i \tau(\mu)} \over \det(\Phi^{v,Ia})} = \int d^4x
d^2\theta  {\Lambda^9 \over \det(\Phi^{v,Ia})}$$ to a non-zero
numerical constant.

\subsubsection{Non-perturbative superpotential for $U(4)$}

As explained in \cite{U(4)} for $U(4)$ gauge theory  with three
chiral multiplets in the {\bf 6}, the D(-1)-D(-1)  geometric
`supermoduli' are $a^{(0)}_\mu$ (instanton position),
$\bar\chi_{I(-2)},\chi^I_{(+2)}$ (internal) and $\Theta^0_{\alpha
(0)}, \bar\Theta_{0\dot\alpha (0)}, \bar\Theta_{I\dot\alpha (-1)}$
(Grassman coordinates), which give $\cN = 1$ superspace measure.
D(-1)-D(3) gauge `supermoduli' are $w^{\dot\alpha}_{u(-1)},
\bar{w}^u_{\dot\alpha (+1)} , \nu^{0}_{u(+1)},
\bar\nu^{0u}_{(-1)}, \nu^{Iu}_{(+1)}$ with $u=1,...4$ $U(4)$ and
$I=1,2,3$ $SU(3)$ `pseudo' flavor respectively.  Notice that the subscript in parentheses represents the charge under
$U(1)_{k_1}$. Taking into account the interactions with D3-D3
excitations $\Phi^{Iuv}=\phi^{Iuv}+ ...$, the fermionic
integration will lead to the determinant
$$\Delta_F = \rho^8 \epsilon^{w_1 w_2 w_3 w_4}\epsilon^{u_1
u_2 u_3 u_4}\epsilon^{v_1 v_2 v_3 v_4} X_{u_1 u_2 v_1 v_2} X_{u_3 u_4 v_3 v_4} Y_{w_1 w_2} Y_{w_3 w_4}$$
with $X=\eps^{IJK} \bar\chi_I \bar\phi_J \bar\phi_K$, $Y_{uv}=\cU_u^{\adot} \cU_{\adot v}$ and $\rho$, $\cU$ are defined
by $w_{u\dot{\alpha}}=\rho \, \cU_{u\dot{\alpha}}$, $\bar{w}^{u\dot{\alpha}}=\rho \, \bar{\cU}^{u\dot{\alpha}}$,
$\bar{\cU}^{u\dot{\alpha}}\cU_{u\dot{\beta}}=\delta^{\dot{\alpha}}_{\dot{\beta}}$ .
Integration over bosonic `moduli' is more involved. For arbitrary choices of the v.e.v's $\bar{\phi}^{Iuv}$
and $\phi^{Iuv}$, even along the flat directions, integration over $\cU$ represents a difficult task.
Fortunately for the choice $\phi^{Iuv}=\eta^{Iuv}$, the full $\phi$-dependence can be factorized. And after
rescaling $\rho^2 \rightarrow \rho^2/(\phi\bar\phi)$, $\chi^I \rightarrow \phi \chi^I$, $\bar{\chi}_I
\rightarrow \bar{\phi} \bar{\chi}_I$, $X_{u_1 u_2 v_1 v_2}\rightarrow
\eps^{I_1 I_2 I_3}\bar{\chi}_{I_1}\bar{\eta}_{I_2 u_1 u_2}\bar{\eta}_{I_3 v_1 v_2}$  the $\phi$-independent
integral $I_B$ becomes
$$I_B = \int d\rho \rho^9 d^{12}\cU d^3\chi d^3\bar{\chi} \Delta _F e^{-\tilde{S}_B}$$
where $\tilde{S}_B = -\rho^2(1 + \eta^{Iuv} Y_{uv} \chi_I + \bar{\eta}_{Iuv} \bar{Y}^{uv}\chi^I +
\bar{\chi}_I\chi^I)$. Restoring the $SU(4)$ gauge and $SU(3)$ 'flavor' invariance the
superpotential follows after promoting $\phi^I \rightarrow \Phi^I $:
$$S_W=c\int d^4a d^2\Theta {\mu^9 e^{2\pi i \tau(\mu)} I_B\over \Phi^6}
= \int d^4x d^2\theta {\Lambda ^9 \over \det_{3\times
3}[\epsilon_{u_1...u_4} \Phi^{I u_1 u_2} \Phi^{J u_3 u_4}] }$$ up
to a non-zero numerical constant.

\subsection{Exotic/Stringy instantons}

EDp-branes on unoccupied nodes of the quiver produce exotic instanton effects. The gauge theory on EDp' is of the same
kind as on EDp (like 8 N-D directions, periodic sector).

Grassmann integration over chiral fermions $\nu$'s at
intersections produces positive powers of $\Phi$. The resulting
non perturbative superpotential can grow at large VEV's, which is
incompatible with field theory intuition (asymptotic freedom) for standard `gauge' instantons.
Yet it is compatible with gauge invariance and `exotic' scaling
$$e^{-A(C')/\ell_s^{p'+1}} \neq e^{-1/g_{YM}^2}$$

For generic $K$, there are many unlifted fermionic zero-modes and one gets
higher derivative F-terms, threshold corrections, ... or
dangerous bosonic zero-modes.
For specific $K$, there are only two unlifted zero-modes ($d^2\theta$) and one gets superpotential terms.
For ED1, the relevant $\nu$'s are in the direction of the worldsheet.

\subsubsection{$U(4)$ model: non-perturbative masses} Let us
consider ``our" $U(4)$ model, $\Theta^0_\alpha$, $X_\mu$ plus 4
$\nu^u$ that couple to one complex component $\phi_{uv}$ (related
to $C$) through
$$ S_{D3-ED3} = \phi_{uv} \nu^u \nu^v + ... $$
The superpotential generated by ED-strings wrapping 2-cycles $C$ passing
through the singularity schematically reads
$$ W(\Phi) = \sum_{C} M_s e^{-A(C)/g_s\ap} \Phi_{C}^2 $$
and thus represents a mass terms for $\Phi \approx A$.


Effect of multi-instanton are hard to evaluate ... Heterotic /
Type I duality may help clarifying the procedure.


\section{Third Lecture: Worldsheet vs D-brane instantons}

\subsection{Heterotic-Type I duality in $D\le 10$}

Perturbatively different string theories may be shown to be equivalent once non-perturbative effects are taken into
account.
Heterotic and Type I string
theories with gauge group $SO(32)$, were conjectured in \cite{WitPol, PWDH}, to be equivalent. In fact, up to field
redefinitions, they share the same low-energy
effective field theory. It was shown in \cite{EW} that for the equivalence to work,
the strong coupling limit of one should correspond to the
weak coupling limit of the other\footnote{Similar situations
in which the strong coupling limit of one string theory is the weak
coupling of another `dual' string theory were discussed earlier by
Duff \cite{Duff}.}. In $D = 10$ the strong - weak coupling duality takes
the following form \cite{WitPol, PWDH}
$$g_s^H = 1/g_s^I,~~~
\alpha^\prime_H = g_s^I
\alpha^\prime_I$$
where $g_s^H$, $\alpha^\prime_H$ and $g_s^I$, $\alpha^\prime_I$ are the heterotic and Type I coupling constants and
tensions respectively. The simple strong-weak coupling duality $\phi_I = -\phi_H$ in $D=10$ changes significantly in lower
dimensions.
Indeed, since the dilaton belongs to the universal sector of the compactification, the relation between the heterotic and
Type I dilatons in $D$ dimensions is determined by dimensional reduction to be \cite{ABPSS}, \cite{ABFPT}
$$\phi_I^{(D)} = {(6-D) \over 4} \phi_H^{(D)} - {(D-2) \over 16} \log\det{G_H^{(10-D)}}$$
where $G_H^{(10-D)}$ is the internal metric in the heterotic-string frame, and there is a crucial sign change at $D=6$
where $\phi_H$ and $\phi_I$ are independent \cite{X}. It is well known that Type I models exist with different number of
tensor multiplets in $D=6$ \cite{MSGP, DPGJ}. This does not have an analogue in perturbative heterotic compactifications
on $K3$. In $D=6$, the Type I dilaton belongs to a hypermultiplet to be identified with one of the moduli of the $K3$
compactification on the heterotic side. In four dimensional $\cN=1$ models on both sides the dilaton appears in a linear
multiplet, and heterotic-type I duality is related to chiral-linear duality. The presence of anomalous $U(1)$'s under
which R-R axions shift suggests that the latter correspond to changed scalars on the heterotic side.

Heterotic-Type I duality requires that the heterotic fundamental string and the Type I D-string be identified. The
massless fluctuations of a Type I D-string are eight bosons and eight negative chirality
fermions in the D1-D1 sector together with 32 positive chirality fermions in the D1-D9 sector. Thus, the world-sheet of
the D-string exactly matches the world-sheet of the Heterotic fundamental string. By the same token, the Type I D5-brane
should be identified with the heterotic NS5-brane. The latter is a soliton of the effective low-energy heterotic action
and its microscopic description is not fully understood. The tensions agree in the two descriptions since
$T_{NS5} = 1/g_H^2 (\ap_H)^3 \equiv 1/g_I (\ap_I)^3 = T_{D5}$.

$SO(32)$ Heterotic / Type I duality has been well tested in $D=10$ and in
toroidal compactifications. In $D=10$ BPS-saturated terms, like $F^4$, $F^2R^2$ and $R^4$, are anomaly related and match
in the two theories as a consequence of supersymmetry and absence of anomaly. In toroidal compactifications, the
comparison of BPS-saturated terms becomes more involved. The spectra of BPS states  become richer and differ on the two
sides at the perturbative level.

Non-perturbative corrections to $F^4$, $F^2R^2$ and $R^4$ terms
are due to instantons that preserve half of the supersymmetry. In
the heterotic string they get perturbative corrections at one loop
only and the NS5-brane is the only relevant non-perturbative
configuration in $D\leq 4$. Instanton configurations can be
provided by taking the world-volume of the NS5-brane to be
Euclidean and to wrap supersymmetrically around a compact
manifold, so as to keep finite the classical action. This requires
at least six-dimensional compact manifold. Therefore,
BPS-saturated terms do not receive non-perturbative corrections
for toroidal compactifications with more than four non-compact
directions. Thus, the full heterotic result arises from tree level
and one loop for $D>4$. In the Type I string both D1- and
D5-branes can provide instanton configurations after
Euclideanization. D5-brane will contribute in four or less
noncompact dimensions, D1-brane can contribute in eight or fewer
noncompact dimensions. Thus, in nine dimensions the two theories
can be compared in perturbation theory. In eight dimensions the
perturbative heterotic result at one-loop corresponds to
perturbative as well as nonperturbative Type I contributions
coming from the D1-instanton via duality. The heterotic results
can be expanded and the Type I instanton terms can be identified.
The classical action can be written straightforwardly and it
matches with the heterotic result. The determinants and multi
instanton summation can also be performed in the Type I theory. In
general, world-sheet instantons in heterotic string duals of Type
I models help clarifying the rules for multi-instanton calculus
with unoriented D-branes. Two prototypical examples are the
$T^4/Z_3$ orbifold to $D=4$, that we have already encountered
\cite{ABPSS}, and the $T^6/Z_2$ orbifold to $D=6$ \cite{MSGP},
that we are going to discuss in the following.

\subsection{Compactification on $T^4/Z_2$ to $D=6$}

\subsubsection{Type I description}

The Type I theory is an un-oriented projection of the Type IIB
theory. Upon compactification on $T^4/Z_2$ to $D=6$, the Type IIB
theory has $\cN =(2,0)$ spacetime supersymmetry with 16
supercharges, \ie those satisfying $Q = R Q$, where $R$ denotes
the inversion of the four coordinates of $T^4$. The $\Omega$
projection preserves only the sum of left- and right-moving
supersymmetries $Q_\alpha +\tilde{Q}_\alpha$. The $\Omega R$
projection preserves the same linear combination since $ Q_\alpha
+ R \tilde{Q}_\alpha \equiv Q_\alpha +\tilde{Q}_\alpha$. The
massless little group in six dimensions is $SO(4)=SU(2)\times
SU(2)$. The massless bosonic content of the unoriented closed
string spectrum contains in untwisted NS-NS sector ({\bf
3,3})+11({\bf 1,1}), in the untwisted R-R sector ({\bf 3,1})+({\bf
1,3})+6({\bf 1,1}), in the twisted NS-NS sector there are 48 ({\bf
1,1}) and in the twisted R-R sector 16 ({\bf 1,1}). This is
exactly the bosonic content of the $D=6$ $\cN=(1,0)$ supergravity
coupled to one tensor and 20 hypermultiplets.

Let us now discuss the unoriented open string spectrum. Tadpole cancellation conditions imply that the total Chan-Paton
dimensionalities of twisted and untwisted sectors both equal to 32. The $U(16)_9\times U(16)_5$ model, which arises at the
maximally symmetric point, where all the D5-branes are on top of an $\Omega$5-plane and no Wilson lines are turned on the
D9-branes, is of particular interest. This model was first discussed by  Bianchi and Sagnotti and later by Gimon and
Polchinski in \cite{MSGP}. The D9-D9 sector contributes a vector multiplet in the adjoint of $U(16)_9$ and hypermultiplets
in the ${\bf 120}_{+2} +{\bf 120}^*_{-2}$. The D5-D5 gives a vector multiplet in the adjoint of $U(16)_5$ and the
hypermultiplet in the ${\bf \tilde{120}}_{+2} +{\bf\tilde{120}}^*_{-2}$.
In the D5-D9 `twisted'\footnote{In the sense that the 4 N-D directions have half-integer bosonic modes.} spectrum there
are half-hypers in the
$({\bf 16}_{+1}, {\bf \tilde{16}}^*_{-1})+({\bf 16}^*_{-1}, {\bf \tilde{16}}_{+1})$ of $U(16)_9\times U(16)_5$.

\subsubsection{Compactification on $T^4/Z_2$ to $D=6$: Heterotic
description}

The Type I model corresponds to a compactification without vector structure \cite{19}, \cite{Berkooz}:
$$\tilde\omega_{2,YM}^{SW} \neq 0 \approx B_2^{NS-NS} = 1/2 (mod~1)$$
where $\tilde\omega_{2,YM}^{SW}$ is modified second Stieffel-Whitney class (obstruction to vector structure).

The $Z_2$ orbifold (besides its geometrical action) acts on the 32 heterotic fermions as
$\lambda_{ws}^A \rightarrow (i) \lambda_{ws}^u,
(-i)\lambda_{ws}^{\bar{u}} $, which breaks the gauge group $SO(32)$ to $U(16)$.
The resulting massless spectrum is as follows.
In the untwisted sector we have four neutral hypers, charged hypers in ${\bf 120}_{+2} +{\bf 120}^*_{-2}$, vector in the
adjoint, one tensor and the $\cN =(1,0)$ supergravity multiplet.

The twisted sector (16 fixed points) does not contain neutral hypermultiplets, it has charged half hypermultiplets in the
${\bf 16}_{-3}+{\bf 16}^*_{+3}$.

\subsubsection{Matching the spectrum}

In order to match the massless spectrums of the two descriptions one has to distribute one `fractional' D5-brane per each
fixed point, thus breaking the D5-brane gauge group $U(16)_5 \rightarrow U(1)_5^{16}$ \cite{Berkooz}.

In six dimensions the full gauge plus gravitational anomaly can be written as \cite{Anomalies & Tadpoles}

$$\cI_8 = \sum_i \left( X^i_2 \wedge X^i_6 +
X^i_4 \wedge \tilde{X}^i_4\right)$$
The GSS counterterm reads as $L_{GSS}=C^{RR}_2 X_4 + \sum_f C^{RR}_{0,f} X^f_6$, so that
Type I photons become massive by eating twisted RR axions:
$\partial C_{0,f}^{RR} \rightarrow D C_{0,f}^{RR} = \partial
C_{0,f}^{RR} + 4 A^{(9)} + A_f^{(5)}$.
The Type I combination $A^I = A^{(9)} - 4 \sum_f A_f^{(5)}$ decouples from twisted
closed string scalars and matches with the heterotic photon $A^{H}$.
The vector multiplets get massive by eating neutral closed string hypers. Thus we have a supersymmetric Higgs-like
mechanism:
full hypers are eaten\footnote{This is an efficient, not fully exploited mechanism for moduli stabilization even in $D=4$.
The remnant of the $D=6$ anomaly in $D=4$ is massive `non-anomalous' $U(1)$'s \cite{Pascal}.}.

\subsection{Duality and dynamics in $D=6$}

In order to further test the correspondence and gain new insights
into multi D-brane instantons, we are going to consider a
four-hyperini Fermi type interaction that is generated by
instantons and corresponds to a `chiral' (1/2 BPS) coupling in the
$\cN=(1,0)$ low energy effective action. If the four hyperini are
localized at four different fixed points, this coupling is absent
to any order in perturbation theory . This is so, because twisted
fields at different fixed points do not interact perturbatively.
ED1-brane or worldsheet instantons which connect the four fixed
points can generate such a term. The contributions will be
exponentially suppressed with the area of the cycle wrapped by the
instanton.

Let us mention what kind of corrections one expects in the two descriptions before describing the computation.
In $D=6$ Heterotic / Type I duality implies
$$\phi_{_H} = \omega_{_I} \qquad , \qquad \phi_{_I} = \omega_{_H} $$
where  $\phi$ is the dilaton and  $\omega$ is the volume modulus.
Supersymmetry implies that there are no neutral couplings between vectors and hypers.
The gauge couplings can only depend (linearly) on the scalar
$\phi_{_H} = \omega_{_I}$ in the unique tensor multiplet, while
$\phi_{_I} = \omega_{_H}$ belongs to a neutral hyper. For these reasons in the heterotic description the hypermultiplet
geometry is tree-level exact, but may get worldsheet instanton corrections
$e^{-h(C)/\alpha^\prime}$, where neutral hypers $h$ determine the size of
2-cycles $C$ in $T^4/Z_2$. In the Type I description, hypers receive both perturbative (string
loops) and non-perturbative corrections from BPS Euclidean
D-string instantons wrapping susy 2-cycles $C$ in $T^4/Z_2$.
The Type I gauge couplings are completely determined by disk amplitudes.
In the heterotic string, they receive (only) a one-loop correction.

\subsection{Four-hyperini amplitude}

\subsubsection{Computational strategy}

Let us summarize our strategy:
\begin{itemize}

\item Focus on a specific 4-hyperini amplitude
$$\cA^{f_1 f_2 f_3 f_4}_{4hyper} = \langle V^{\zeta, f_1}_{\bf 16}V^{\zeta, f_2}_{{\bf 16}^*} V^{\zeta,
f_3}_{\bf 16} V^{\zeta, f_4}_{{\bf 16}^*} \rangle $$ absent at tree level for
particular choices of fixed points

\item Compute $\cA^{f_1 f_2 f_3 f_4}_{4hyper}$ in the limit of
vanishing momenta

\item Start with heterotic string, where it is tree level exact and
extract worldsheet instanton corrections

\item Translate into Type I language and interpret the result in
terms of perturbative and non-perturbative contributions

\item Learn new rules for unoriented multi D-brane instantons

\end{itemize}

\subsubsection{Heterotic description}

To compute the four-hyperini Fermi interaction in the heterotic description we need the hyperini vertex operators
$$ V^\zeta_{{\bf 16}/{\bf
16}^*} = \zeta^{u/\bar{u}}_{f,a}(p) S^a e^{-\varphi/2}(z)
\tilde\Sigma^{{u/\bar{u}}}(\bar{z}) \sigma_f e^{ipX}(z,\bar{z})$$
where $\sigma_f$ is the bosonic $Z_2$-twist field ($h= 1/4$),
$\tilde\Sigma^{{u/\bar{u}}} = :e^{\pm i \tilde\phi_u}\prod_v
e^{\mp i \tilde\phi_v/4}:$ are twisted ground-states ($h=3/4$) for
heterotic fermions $\tilde\lambda^{u/\bar{u}}$, $S^a$ are $SO(5,1)$ spin fields, $\varphi$ and $\tilde\phi_u$ are the
bosonizations of the superghost and $SO(32)$ gauge fermions respectively.
One can use $SL(2,C)$ invariance on the sphere to set $z_1\rightarrow \infty$,
$z_2\rightarrow 1$, $z_3\rightarrow z$, $z_4\rightarrow 0$ with
cross ratio $z = z_{12} z_{34} / z_{13} z_{24}$. Then the string amplitude will depend on the $SL(2,C)$ invariant cross
ratio $z$.

The $Z_2$-twist field correlator is given by \cite{DFMS}
$$ \langle \prod_{i=1}^4 \, \sigma_{f_i} (z_i,\bar{z}_i)
 \rangle \rightarrow
 |z_\infty|^{-1} \,\Psi_{qu}(z,\bar z)  \,\Lambda_{cl}
 \left[^{\vec{f}_{12}}
 _{\vec{f}_{13}}
 \right](z,\bar z)\label{z4twist} $$
The quantum part $\Psi_{qu}$ is independent of the twist-fields locations \ie
of the choice of 4 out 16 fixed points $\vec{f}_i= 1/2(\epsilon^1_i,
\epsilon_i^2, \epsilon_i^3, \epsilon_i^4)$ with $\epsilon_i^a=
0,1$ and in order to get a non-trivial coupling the $\vec{f}_i$ should satisfy $\sum_i \vec{f}_i=\vec{0}$  mod
$\Lambda(T^4)$.
$\Lambda_{cl}=\sum e^{-S_{inst}}$ is the classical part accounting for worldsheet
instantons depending on the ralative positions $\vec{f}_{ij}=\vec{f}_i-\vec{f}_j$.
The $Z_2$-twist field correlator can be mapped to the torus doubly covering the sphere with two $Z_2$ branch cuts using
the relation between the cross-ratio $z$ and the Teichm\"uller parameter of the torus $\tau (z)$
$$ z = {\vartheta_3^4(\tau) /\vartheta_4^4 (\tau) } .$$
The quantum and classical parts of the 4-twist correlator read
$$ \Psi_{qu} (z,\bar z)= 2^{-{8/3}}\,  |z(1-z)|^{-{1/3}}\, \tau_2^{-2}|\eta(\tau)|^{-8} $$
$$ \Lambda_{cl} \left[^{\vec{f}_{12}}_{\vec{f}_{13}}
\right](z,\bar z)=  \sum_{\vec{m}, \vec{n}} e^{ -{\pi\over\tau_2(z)} ( \vec{m}+\vec{n}\tau + \vec{f}_{13}+
\vec{f}_{12}\tau ) \cdot (G+B)\cdot ( \vec{m}+\vec{n}\bar\tau + \vec{f}_{13}+ \vec{f}_{12}\bar\tau )} $$
where $G_{ij}$ is the metric and $B_{ij}$ is the antisymmetric tensor of $T^4/Z_2$
(neutral hypers).
Writing the $z$-integral as integral over the torus modulus $\tau$ (for
$s,t\to 0$) one finds
$$ {\cal A}^{f_1,f_2,f_3,f_4}_{u_1 \bar u_2  u_3 \bar{u}_4}
= \cV(T^4) \int_{\cF_2}  {d^2\tau \over \tau_2^2}
\left( {\bar \vartheta_4^4 \over \bar\vartheta_3^4 } \delta_{u_1 \bar{u}_2} \delta_{u_3 \bar u_4}
- {\bar \vartheta_4^4 \over \bar\vartheta_2^4 }\,
\delta_{u_1 \bar{u}_4} \delta_{u_3 \bar u_2}\right)\,
\Lambda_{cl} \left[^{\vec{f}_{12}}_{\vec{f}_{13}}\right]. $$
The integral goes over the fundamental domain $\cF_2$ of the index 6 subgroup $\Gamma_2$ of
$SL(2,Z)$, leaving invariant $\vartheta_{even}$ \cite{KM}.
The region $\cF_2$ can be decomposed into 6 domains each of which is an image of the fundamental domain $\cF$  of $SL(2,
{Z})$ under the action of the 6 elements of $SL(2,{Z})/\Gamma_2$
$$\int_{\cF_2} {d^2\tau\over \tau_2^2} \, \Phi(\tau,\bar{\tau})=
  \int_{\cF} {d^2\tau\over \tau_2^2} \, \sum_{s=1}^6 \, \Phi(\tau_s,\bar{\tau}_s)\label{f2f} $$
where $\tau_s=\gamma_s(\tau)$, $\gamma_s =\{1,S,T,TS,ST,TST\}$.
For the 4-hyperini amplitude one gets
$$ \Phi(\tau,\bar{\tau})=  \left( {\bar \vartheta_4^4 \over \bar\vartheta_3^4 } \delta_{u_1 \bar{u}_2} \delta_{u_3 \bar
u_4} -{\bar \vartheta_4^4 \over \bar\vartheta_2^4 }\,\delta_{u_1 \bar{u}_4} \delta_{u_2 \bar u_3}\right)\,
 \Lambda_{cl} \left[^{\vec{f}_{12}} _{\vec{f}_{13}} \right] .$$

In the special case when all 4-hyperini are located at the same fixed point $\vec{f}_{12}=\vec{f}_{13}=(\vec{0})$, the
amplitudes receive contribution only from BPS-like modes as in Type I (see later). The instanton sum
$\Lambda_{cl}\left[^{\vec{0}}_{\vec{0}}\right]$ is modular invariant. Sums over 6 images produce
$$ \sum_{s=1}^6 \, {\bar \vartheta_4^4 \over \bar\vartheta_3^4 } (\bar{\tau}_s)=3
 \quad , \quad
\sum_{s=1}^6 {\bar \vartheta_4^4 \over \bar\vartheta_2^4 } (\bar{\tau}_s)= -3 $$
and the final expression for the amplitude with $\vec{f}_1= \vec{f}_2 = \vec{f}_3 = \vec{f}_4$ is given by
$$ {\cal A}^{f_1,f_1,f_1,f_1}_{u_1 \bar u_2  u_3 \bar{u}_4} = 3 (\delta_{u_1 \bar{u}_2} \delta_{u_3 \bar u_4}
+\delta_{u_1 \bar{u}_4} \delta_{\bar{u}_2 u_3})\, \cV(T^4) \int_{\cF} {d^2\tau\over \tau_2^2} \,
\Lambda_{cl} \left[^{\vec{0}}_{\vec{0}} \right]. $$
Next consider the case when hyperini are located in pairs at two different fixed point:
\begin{itemize}
\item $\vec{f}_{12}=\vec{f}_{13} = \vec{f}$ for $ \delta_{u_1 \bar{u}_2}
\delta_{u_3 \bar u_4}$ structure in 4-hyperini amplitude
$$ {\cal A}^{f_1,f_2,f_2,f_1}_{u_1 \bar u_1  u_3 \bar{u}_3}
 =\cV(T^4) \int_{\cF} {d^2\tau\over \tau_2^2} \, \left( \Lambda_{cl}
\left[^{\vec{f}}_{\vec{f}}\right] +\Lambda_{cl}\left[^{\vec{f}}_{\vec{0}}\right]
 +\Lambda_{cl} \left[^{\vec{0}}_{\vec{f}}\right] \right). $$
\item $\vec{f}_{12}=\vec{0}, \vec{f}_{13}=\vec{h}$ for
$\delta_{u_1 \bar{u}_4} \delta_{u_3 \bar u_2}$ structure in
4-hyperini amplitude
$${\cal A}^{f_1,f_1,f_3,f_3}_{u_1 \bar u_2  u_2 \bar{u}_1} = \cV(T^4)
\int_{\cF} {d^2\tau\over \tau_2^2} \, \left( \Lambda_{cl}
\left[^{\vec{0}}_{\vec{h}}\right] +\Lambda_{cl}
\left[^{\vec{h}}_{\vec{0}}\right]
+\Lambda_{cl} \left[^{\vec{h}}_{\vec{h}}\right]\right). $$
\end{itemize}
These are the same integrals as for BPS saturated thresholds to $F^4$ in $T^4$
compactifications (with shifts) \cite{KM}. Since the pieces proportional to $\delta_{u_1 \bar{u}_2} \delta_{u_3 \bar u_4}$
and $\delta_{u_1 \bar{u}_4} \delta_{u_3 \bar u_2}$ are related by a simple relabeling of the fixed points $f_i$'s, we have
restricted our attention onto the amplitude with color structure $\delta_{u_1 \bar{u}_2} \delta_{u_3 \bar u_4}$ for the
first and the amplitude with color structure $\delta_{u_1 \bar{u}_4} \delta_{u_3 \bar u_2}$ for the second case.
Performing Poiss\`on resummation over $\vec{m}$ in $\Lambda _{cl}$ one finds
$$\Lambda_{cl} \left[^{\vec{0}}_{\vec{f}}\right] = {\tau_2^2 \over
\cV(T^4)} \sum_{\vec{k}, \vec{n}} (-)^{2\vec{f}\cdot\vec{k}}
q^{p_L^2/2} \bar{q}^{p_R^2/2}$$
where $\vec{p}_{L/R} = {1\over \sqrt{2}} (E^{-1}\vec{k} + E^t
\vec{n}) $ , $E E^t = G$ and we have set $B=0$ for simplicity. One can recognize the shifted orbifold partition function
$$ \Lambda_{cl} \left[^{\vec{0}}_{\vec{f}}\right] (G) +\Lambda_{cl}\left[^{\vec{f}}_{\vec{0}}\right](G)
+\Lambda_{cl} \left[^{\vec{f}}_{\vec{f}}\right](G)=2 \Lambda_{cl} \left[^{\vec{0}}_{\vec{0}}\right] (G_{(\vec{f})}) -
\Lambda_{cl} \left[^{\vec{0}}_{\vec{0}}\right] (G).$$
The toroidal metric $G_{(\vec{f})}$ is `halved' along the direction
$\vec{v} = 2\vec{f}$ by $SO(d,d)$ transformation.
This is similar to what one gets for the threshold corrections to $F^4$ terms in toroidal compactifications, reviewed in
Sect. 5.1.

\subsubsection{Type I description}

As we already noticed, in the Type I description hypers can receive both perturbative and non-perturbative corrections
since the dilaton belongs to a hypermultiplet. Some scattering amplitudes may vanish in perturbation theory and receive
only contributions from non-perturbative effects. The four-hyperini Fermi interaction term does not get perturbative
contributions for $f_i$ all different from one another or for $f_1=f_3$ (same charge). When all fixed points $f_i$ are
equal or are equal in pairs $f_1=f_2$ or $f_2=f_3$ (opposite charge) there is a perturbative correction which matches with
the contribution of the degenarate orbit in the heterotic description.

The open string vertex operators are given by
$$ V^\zeta_{\bf 16} = \zeta^f_a(p) S^a e^{-\varphi/2}
\sigma_f e^{ipX} \Lambda^u_f \qquad V^\zeta_{{\bf 16}^*} =
\zeta^f_a(p) S^a e^{-\varphi/2} \sigma_f e^{ipX} \bar\Lambda^f_{u}$$
They involve Chan-Paton matrices $\Lambda ^u_f$ in the bifundamental of $U(16)\times U(1)_5^{16}$ rather than heterotic
fermions $\lambda$ yielding
$$ Tr(\Lambda^{u_1}_{f_1} \Lambda^{\bar{u}_2}_{f_2}
\Lambda^{u_3}_{f_3}\Lambda^{\bar{u}_4}_{f_4}) = \delta_{f_1f_2}
\delta_{f_3f_4} \delta^{u_1\bar{u}_4} \delta^{u_3\bar{u}_2} +
\delta_{f_1f_4} \delta_{f_3f_2} \delta^{u_1\bar{u}_2}
\delta^{u_3\bar{u}_4}. $$

Consider $Z_2$-twist field correlator for open strings with 4 N-D boundary
conditions \cite{CPKW}.
The quantum part of the 4-twist correlator, which is independent of location $f_i$ of twist fields, is given by
$$ \Psi_{qu} = [x(1-x)]^{-1/3} t(x)^{-2} \eta(it)^{-4} $$
where $ x ={\vartheta_3^4(it) / \vartheta_4^4 (it) }$ is $SL(2,R)$ invariant ratio with $t$ the modular parameter of the
annulus doubly covering the disk.
The classical part from exchange of (massive) open string modes
stretched between (different) fixed points is
$$\Lambda_{cl} [^{f_{12}}_{f_{13}}]= \delta^{f_1}_{f_2} \delta^{f_3}_{f_4}
\sum_{\vec{n}} e^{- {\pi t} (\vec{n}+\vec{f}_{14})^t G
(\vec{n}+\vec{f}_{14})} + \delta^{f_1}_{f_4} \delta^{f_3}_{f_2}
\sum_{\vec{n}} e^{- {\pi t} (\vec{n}+\vec{f}_{12})^t G
(\vec{n}+\vec{f}_{12})}.$$
Plugging into the open string amplitude and taking the limit $s,t\rightarrow 0$ one finds a perfect agreement with
heterotic degenerate orbits, which is independent of $B_2^H \approx C_2^{R-R}$ but only on $G$ and $\phi$
(recall $\omega_H = \phi_I$, $\phi_H = \omega_I$). Terms involving $B_2^H$  have no disk counterpart in the Type I
description since the dual $C_2^{RR}$ couples to (E)D-strings.

\subsubsection{ED-string corrections}

We then consider non-perturbative corrections to the four-hyperini coupling in the Type I description.
When the four hyperini are located at the different fixed points $f_i$ we have only non
perturbative contribution from (regular) ED-strings wrapping
supersymmetric (untwisted) two cycles $C \approx T^2/Z_2 = S^2$,
passing through the four fixed points.
Fractional ED-strings wrapping the 16 collapsed `rigid'
2-cycles (since corresponding moduli are eaten by anomalous
$U(1)_{5}^{16}$) may contribute to amplitudes having also
perturbative contributions.

Using by now the well established Heterotic / Type I duality we will deduce
the `exact' 4-hyperini amplitude and determine
ED-string corrections. Then we interpret these corrections in terms of symmetric orbifold CFT \cite{GMB}.

Let us describe the spectrum of ED-strings. The instanton dynamics is governed by a gauge theory describing the
excitations of unoriented strings connecting E1, D5 and D9 branes. Three sectors of open string excitations are
\begin{itemize}
\item E1-E1 strings (2 N-N, 8 D-D): $X^I$, $S^a$,
$\tilde{S}^{\dot{a}}$ with $I=1,...,8_v$, $a, \dot{a}=1,...,4$
\item E1-D9 strings (2 N-N, 8 N-D):
$\lambda^u, \lambda_{\bar{u}}$ with $u, \bar{u} = 1, ..., 16$
\item E1-D5 strings (2 D-D, 8 N-D): $\mu^f, \mu_{\bar{f}}$ with $f=f_1, f_2, f_3, f_4$
\end{itemize}

Alternatively, after T-duality along the wrapped 2-cycle we will have $E1
\rightarrow E(-1)$, $D9 \rightarrow D7_9$, $D5 \rightarrow D7_5$.
The residual (super)symmetry of the spectrum is
$$\cN =(8,0) \rightarrow \cN = (4,0).$$
And the spacetime symmetry breaks according to
$$SO(9,1) \rightarrow SO(5,1)\times SU(2) \times SU(2) \rightarrow SO(5,1) \times SO(2)_E\times SO(2).$$
$\cN = (4,0)$ gauge theory in IR flows to symmetric product CFT
$$({ R}^6\times T^4/Z_2)^k/S_k.$$
ED-string wraps two-cycle $C$ inside $T^4/Z_2$ which is
specified by the two vectors $M_k=(\vec{k}_1,\vec{k}_2)$ each made out of four integers with greatest common divisor 1.
$\vec{k}_{1,2}$ show how many times the two 1-cycles of $C$ wrap around 1-cycle of $T^4/Z_2$.

Heterotic vertex operator can be derived from interaction term with hyperino:
$$\cL_{4F} = (\zeta_a)^u_f \mu^f_i \lambda^i_u S^a = V^H_\zeta$$
Only $(\ell)^m$-twisted sectors (with $m\ell=k$ and ${Z}_{\ell}^s$
projection) with exactly four fermionic zero modes of $S^a$
contribute. So, we can fold the $k$ copies of fields and form a single field on a worldsheet with the following Kahler and
complex structures:
$$\cT(M)=k \,\cT(M_k) \quad , \quad \cU(M)={m\,\cU(M_k)+s \over \ell}$$
where $M=M_k \left( \begin{array}{cc}
l & s \\
0 & m  \\
\end{array} \right) $.
This is in perfect agreement with the heterotic result $\cI_{d,d}^{ndeg}$ for the four-hyperino coupling on $T^4/Z_2$.

This is the generalized Hecke transform as in the Heterotic computation!

\subsection{Summary and outlook}

Let us summarize the content of our lectures
\begin{itemize}
\item {There are two classes of unoriented D-brane instantons:}
\begin{itemize}

\item{ `Gauge instantons' may generate a VY-ADS-like superpotential of the form
$$W \approx {\Lambda^{\beta} \over \phi^{\beta -3}}$$
where $\beta$ is the one-loop coefficient in the expansion of the $\beta$ function and
$\Lambda^{\beta} = M_s^\beta e^{-T(C)}$.}

\item {`Exotic instantons' may generate a non-perturbative
superpotential of the form $$W \approx M_s^{3-n} e^{-S_{EDp'}(C')} \phi^n
\quad (n=0,1,...)$$

The thumb rule is the existence of exactly two unlifted fermionic zero-modes. We illustrated this situation with
$T^6/Z_3$.}
\end{itemize}
\item{Combining the two kinds of superpotentials one can achieve (partial) moduli stabilization and SUSY breaking! The
same may happen when only one kind of superpotential is generated in the presence of fluxes and other dynamical effects,
such as FI terms \cite{MBJFMFF}...}

\item{When extra zero-modes are present, threshold corrections to (higher-derivative) terms may arise. We illustrated this
possibility for a compactification to $D=6$ on $T^6/Z_2$, where a fully non-perturbative four hyperini amplitude (Fermi
interaction) can be computed exploiting Heterotic - Type I duality. }

Threshold corrections to gauge couplings in freely acting orbifolds $T^6/Z_2\times Z_2$
were computing by similar means.

\item A  by-product of the analysis in $D=6$, an economical mechanism of moduli
stabilization can be exploited whereby non-anomalous $U(1)$'s in $D=4$ eat would-be hypers due to anomalies in $D=6$.

\item The behaviour of D-brane instanton effects in the presence of fluxes or under wall crossing and the reformulation of
(unoriented) D-brane instanton calculus in terms of localization are extremely active subjects. We hope the interested
reader could consult part of the vast literature on the subject \cite{Lerdaetal, FMP, Blum, BWnieist, instflux, U(4),
MBJFMFF, BCKWreview}.
\end{itemize}

There is a long way to go ... and a lot to learn on
unoriented D-brane instanton.

\vskip 1cm \noindent {\large {\bf Acknowledgments}} \vskip 0.2cm
We would like to thank M.~Bill\`o, M.~Cvetic, E.~Dudas, D.~Krefl,
A.~Lerda, S.~Sethi, A.~Uranga, T.~Weigand for useful discussions
and especially F.~Fucito, E.~Kiritsis, S.~Kovacs, J.~F.~Morales,
R.~Poghossyan, G.~Pradisi, G.C.~Rossi for fruitful collaboration
on topics related to these Lectures. We would also like to thank
C.~Kounnas and N.~Toumbas for organizing the Fourth  Young
Researchers Workshop {\it ``Strings @ Cyprus''}, September 2008,
and creating a very stimulating atmosphere in Kounnas Bay. M.~B.
would like to thank NITheP for the kind hospitality at STIAS,
Stellenbosch, South Africa during completion of these notes. This
work was partially supported by the The European Superstring
Theory Network MRTN-CT-2004-512194, by the ERC Advanced Grant
n.226455 {\it ``Superfields''} and by the Italian MIUR-PRIN
contract 20075ATT78 {\it ``Symmetries of the Universe and of the
Fundamental Interactions''}.



\begin{thebibliography}{99}

\bibitem{BCKWreview}
  M.~Frau and A.~Lerda,
  Fortsch.\ Phys.\  {\bf 52}, 606 (2004)
  [arXiv:hep-th/0401062].
  A.~M.~Uranga,
  JHEP {\bf 0901}, 048 (2009)
  [arXiv:0808.2918 [hep-th]].
Instantons in Type II String Theory'', Invited review to appear in
Annu.Rev.Nuc.Part.Sci (59) 2009, arXiv:0902.3251;
 M. Cvetic,
``D-brane Instantons'', Lectures delivered at the School ``New
Perspectives in String Theory'', GGI Arcetri (Florence), June
2009.

\bibitem{Kirbook}
  E.~Kiritsis,
{\it  Princeton, USA: Univ. Pr. (2007) 588 p}
  K.~Becker, M.~Becker and J.~H.~Schwarz,
{\it  Cambridge, UK: Cambridge Univ. Pr. (2007) 739 p}

\bibitem{ADHM}
M. Atiyah, V. Drinfeld, N. Hitchin and Yu. Manin, " Construction of Instantons", Phys. Lett. {\bf A 65} (185) 1978.

\bibitem{MKR}
M. Bianchi, S. Kovacs, G. Rossi, "Instantons and Supersymmetry" Lect.NotesPhys. {\bf 737} (303) 2008, arXiv:hep-
th/0703142.

\bibitem{Douglas'95}
M. R. Douglas, "Branes within Branes " arXiv:hep-th/9512077.
  E.~Witten,
  Nucl.\ Phys.\  B {\bf 460}, 335 (1996)
  [arXiv:hep-th/9510135].

\bibitem{Dine}
M. Dine, N. Seiberg, X. G. Wen and E. Witten, "Non-perturbative effects on the string world-sheet", Nucl. Phys. B {\bf
278}, (769) 1986;

M. Dine, N. Seiberg, X. G. Wen and E. Witten, "Non-perturbative effects on the string world-sheet. 2", Nucl. Phys. B {\bf
289}, (319) 1987.

\bibitem{DualitInst}
  A.~Sen,
  arXiv:hep-th/9802051.
  E.~Kiritsis,
  arXiv:hep-th/9906018.

\bibitem{25}
L. J Dixon, V. Kaplunovski and J. Louis, "Moduli dependence of string loop corrections to gauge coupling constants", Nucl.
Phys. B {\bf 355}, (649) 1991;

I. Antoniadis, E. Gava and K. S. Narain, "Moduli corrections to gauge and gravitational couplings in four-dimensioanl
superstrings", Nucl. Phys. B {\bf 383} (93) 1992, arxiv:hep-th/9204030;

E. Kiritsis and C. Kounnas, "Infrared regularization of superstring theory and the one loop calculation of coupling
constants", Nucl. Phys. B {\bf 442} (472) 1995, arxiv:hep-th/9501020; Nucl. Phys. Proc. Suppl. {\bf 41} (1995) 331,
arxiv:hep-th/9410212.

\bibitem{Becker}
K. Becker, M. Becker and A. Strominger, "Five-branes, membranes and nonperturbative string theory", Nucl. Phys. B {\bf
456}, (130) 1995, arxiv:hep-th/9507158.

\bibitem{Billo}
M. Billo, M. Frau, I. Pesando, F. Fucito, A. Lerda and A. Liccardo, "Classical gauge instantons from open strings", JHEP
{\bf 0302}, (045) 2003, arxiv:hep-th/0211250;

M.Billo, M. Frau, F. Fucito and A. Lerda, "Instanton calculus in R-R background and the topological string", JHEP {\bf
0611}, (012) 2006, arxiv:hep-th/0606013.

\bibitem{Dorey}
N. Dorey, T.J. Hollowood, V.V. Khoze, M.P. Mattis and S. Vandoren, " Multi-Instanton Calculus and the AdS/CFT
Correspondence in N=4 Superconformal Field Theory", Nucl. Phys. B{\bf 552} (88) 1999, arXiv:hep-th/9901128.

\bibitem{BKB}
C. Bachas, C. Fabre, E. Kiritsis, N. A. Obers and P. Vanhove, "Heterotic/type-I duality and D-brane instantons", Nucl.
Phys. B {\bf 509} (33) 1998, arxiv:hep-th/9707126;

E. Kiritsis and N. A. Obers, "Heterotic/type-I duality in $D<10$ dimensions, threshold corrections and D-instantons", JHEP
{\bf 9710} (004) 1997, arxiv:hep-th/9709058;

C. Bachas, "Heterotoc versus Type I", Nucl. Phys. Proc. Suppl. {\bf 68} (348) 1998, arxiv:hep-th/9710102.

\bibitem{WitPol}
J. Polchinski, E. Witten, "Evidence for Heterotic - Type I String Duality", Nucl. Phys. B {\bf 460} (525) 1996, arXiv:hep-
th/9510169;

\bibitem{Kiritsis}
L. J. Dixon, V. Kaplunovsky, J. Louis, "On Effective Field Theories Describing (2,2) Vacua of the Heterotic String", Apr
1989, Nucl. Phys. B {\bf 329} (27) 1990;

C. Bachas, C. Fabre, E. Kiritsis, N. A. Obers, P. Vanhove, "Heterotic / Type I duality and D-brane instantons", Nucl.
Phys. B {\bf 509}, (33) 1998, arXiv:hep-th/9707126;

W. Lerche, S. Stieberger, "1/4 BPS States and Non-Perturbative Couplings in N=4 String Theories",  Adv. Theor. Math. Phys.
{\bf 3}, (1539) 1999, arXiv:hep-th/9907133.

\bibitem{MBEGJFMKN}
M. Bianchi, E. Gava, J. F. Morales, K. S. Narain, "D-strings in unconventional type I vacuum configurations",
Nucl.Phys. B {\bf 547} (96) 1999, arXiv:hep-th/9811013.

\bibitem{Gutperetal}
M. Gutperle, "A note on heterotic/type I' duality and D0 brane quantum mechanics", JHEP {\bf 9905} (007) 1999,
arXiv:hep-th/9903010.

\bibitem{KFSS}
K. Foerger, S. Stieberger, "Higher Derivative Couplings and Heterotic-Type I Duality in Eight Dimensions", Nucl.Phys.
B{\bf 559}  (277) 1999, arXiv:hep-th/9901020.

\bibitem{Lerdaetal}

M. Billo', L. Ferro, M. Frau, L. Gallot, A. Lerda, I. Pesando, "Exotic instanton counting and heterotic/type I'
duality", JHEP {\bf 0907} (092) 2009, arXiv:0905.4586;

\bibitem{FMP}
F. Fucito, J. F. Morales, R. Poghossian, "Exotic prepotentials from D(-1)D7 dynamics", arXiv:0906.3802.

\bibitem{ABDFPT}
I. Antoniadis, C. Bachas, E. Dudas, "Gauge couplings in four-dimensional Type I string orbifolds", Nucl.Phys. B{\bf 560}
(93) 1999, arXiv:hep-th/9906039.

I. Antoniadis, C. Bachas, C. Fabre, H. Partouche, T.R. Taylor, "Aspects of Type I - Type II - Heterotic Triality in Four
Dimensions", Nucl.Phys. B{\bf 489} (160) 1997, arXiv:hep-th/9608012.

\bibitem{Blum}
R. Blumenhagen, M. Cvetic and T. Weigand, "Spacetime instanton corrections in $4D$ string vacua - the seesaw mechanism
for D-brane models", Nucl. Phys .B {\bf 771} (113) 2007, arxiv:hep-th/0609191;

M. Haack, D. Krefl, D. Lust, A. Van Proeyen and M. Zagermann, "Gaugino condensates and $D$-terms from D7-branes", JHEP
{\bf 0701} (078) 2007, arxiv:hep-th/0609211;

L. E. Ibanez and A. M. Uranga, "Neutrino Majorana masses from string theory instanton effects", JHEP {\bf 0703} (052)
2007, arxiv:hep-th/0609213;

B. Florea, S. Kachru, J. McGreevy and N. Saulina, "Stringy instantons and quiver gauge theories", JHEP {\bf 0705} (024)
2007, arxiv:hep-th/0610003.

\bibitem{Silverstein}
E. Silverstein and E. Witten, "Criteria for conformal invariance of (0,2) models", Nucl. Phys. B {\bf 444}, (161) 1995,
arxiv:hep-th/9503212;

\bibitem{BWnieist}
C. Beasley and E, Witten, "New instanton effects in string theory", JHEP {\bf 0602} (060) 2006, arxiv:hep-th/0512039.

\bibitem{FW}
D. S. Freed and E. Witten, "Anomalies in string theory with D-branes", arxiv:hep-th/9907189.

\bibitem{KPT}
A. K. Kashani-Poor and A. Tomasiello, "A stringy test of flux-induced isometry gauging", Nucl. Phys. B {\bf 728}, (135)
2005, arxiv:hep-th/0505208.

\bibitem{instflux}
M. Billo', L. Ferro, M. Frau, F. Fucito, A. Lerda, J. F. Morales, "Flux interactions on D-branes and instantons",
JHEP {\bf 0810} (112) 2008, arXiv:0807.1666.

\bibitem{U(4)}
M. Bianchi, F. Fucito, J. F. Morales, "D-brane Instantons on the $T^6/Z_3$ orientifold", JHEP{\bf 0707} (038) 2007,
arxiv:hep-th/0704.0784.

\bibitem{Anom&Tad}
M. Bianchi, J. F. Morales, "Anomalies and Tadpoles", JHEP {\bf 0003} (030) 2000, arXiv:hep-th/0002149.

\bibitem{AIMU}
G. Aldazabal, D. Badagnani, L.E. Ibanez, A.M. Uranga, "Tadpole versus anomaly cancellation in D=4,6 compact IIB
orientifolds" JHEP {\bf 9906} (031) 1999, arXiv:hep-th/9904071.


\bibitem{Lionettoetal}
P. Anastasopoulos, F. Fucito, A. Lionetto, G. Pradisi, A. Racioppi, Y. S. Stanev, "Minimal Anomalous U(1)' Extension of
the MSSM", Phys.Rev.{\bf D78}, (085014) 2008, arXiv:0804.1156 ,

F. Fucito, A. Lionetto, A. Mammarella, A. Racioppi, "Axino Dark Matter in Anomalous U(1)' Models", arXiv:0811.1953.

\bibitem{PWDH}
A. Dabholkar, "Ten Dimensional Heterotic String as a Soliton", Phys. Lett. B {\bf 357} (307) 1995, arXiv:hep-th/9506160;

C. M. Hull, "String-String Duality in Ten Dimensions", Phys. Lett. B {\bf 357} (545) 1995, arXiv:hep-th/9506194.

\bibitem{EW}
E. Witten, "String Theory Dynamics In Various Dimensions", Nucl. Phys. B {\bf 443} (85) 1995, arXiv:hep-th/9503124.

\bibitem{ABPSS}
C. Angelantonj, M. Bianchi, G. Pradisi, A. Sagnotti, Y. Stanev, "Chiral Asymmetry in Four-Dimensional Open-String Vacua",
Phys. Lett. B {\bf 385} (96) 1996, arXiv:hep-th/9606169.

\bibitem{ABFPT}
I. Antoniadis, C. Bachas, C. Fabre, H. Partouche, T.R. Taylor, "Aspects of Type I - Type II - Heterotic Triality in Four
Dimensions" Nucl. Phys. B {\bf 489} (160) 1997, arXiv:hep-th/9608012 .

\bibitem{X}
M. Berkooz, R. G. Leigh, J. Polchinski, J. H. Schwarz, N. Seiberg, E. Witten, "Anomalies, Dualities, and Topology of D=6
N=1 Superstring Vacua", Nucl. Phys. B {\bf 475} (115) 1996, arXiv:hep-th/9605184.

\bibitem{MSGP}
M. Bianchi, A. Sagnotti, "Twist symmetry and open string Wilson lines", Nucl. Phys. B {\bf 361} 519 (1991),
"On the systematics of open string theories", Phys. Lett. B {\bf 247} (517) 1990;

E. G. Gimon, J. Polchinski, "Consistency Conditions for Orientifolds and D-Manifolds", Phys. Rev. D {\bf 54} (1667) 1996,
arXiv:hep-th/9601038.

\bibitem{DPGJ}
A. Dabholkar, J. Park, "Strings on Orientifolds", Nucl. Phys. B {\bf 477} (701) 1996, arXiv:hep-th/9604178 and "An
Orientifold of Type-IIB Theory on $K3$", Nucl. Phys. B {\bf 472} 207 (1996), arXiv:hep-th/9602030.

E. G. Gimon, C. V. Johnson, "$K3$ Orientifolds", Nucl. Phys. B {\bf 477} (715) 1996, arXiv:hep-th/9604129.

\bibitem{Duff}
M. J. Duff, "Strong/Weak Coupling Duality from the Dual String", Nucl. Phys. B {\bf 442} (47) 1995, arXiv:hep-th/9501030.

\bibitem{19}
M. Bianchi, G. Pradisi, A. Sagnotti, "Toroidal compactification and symmetry breaking in open string theories", Nucl.
Phys. B {\bf 376} (365) 1992;

Z. Kakushadze, "Aspects of N=1 Type I-Heterotic Duality in Four Dimensions", Nucl. Phys. B {\bf 512} (221) 1998,
arXiv:hep-th/9704059;

M. Bianchi,"A Note on Toroidal Compactifications of the Type I Superstring and Other Superstring Vacuum Configurations
with 16 Supercharges", Nucl. Phys. B {\bf 528} (73) 1998, arXiv:hep-th/9711201;

E. Witten, "Toroidal Compactification Without Vector Structure", JHEP {\bf 9802} (006) 1998, arXiv:hep-th/9712028;

C. Angelantonj and R. Blumenhagen, "Discrete Deformations in Type I Vacua", Phys. Lett. B {\bf 473} (86) 2000, arXiv:hep-
th/9911190.

\bibitem{Berkooz}
M. Berkooz, R. G. Leigh, J. Polchinski, J. H. Schwarz, N. Seiberg, E. Witten, "Anomalies, Dualities, and Topology of D=6
N=1 Superstring Vacua", Nucl. Phys. B {\bf 475} (115) 1996, arXiv:hep-th/9605184.

\bibitem{Anomalies & Tadpoles}
M. B. Green, J. H. Schwarz and P. C. West, "Anomaly Free Chiral Theories in Six-Dimensions", Nucl. Phys. B {\bf 254} (327)
1985;

A. Sagnotti, "A Note on the Green - Schwarz Mechanism in Open - String Theories", Phys. Lett. B {\bf 294} (196) 1992,
arXiv:hep-th/9210127.

\bibitem{Pascal}
P. Anastasopoulos, "Anomalous U(1)s masses in non-supersymmetric open string vacua", Phys. Lett. B {\bf 588} (119) 2004,
arXiv:hep-th/0402105;

P. Anastasopoulos, M. Bianchi, E. Dudas, E. Kiritsis, "Anomalies, Anomalous U(1)'s and generalized Chern-Simons terms",
JHEP {\bf 0611} (057) 2006, arXiv:hep-th/0605225.

\bibitem{DFMS}
L. J. Dixon, D. Friedan, E. J. Martinec, S. H. Shenker, " The Conformal Field Theory of Orbifolds", Nucl. Phys. B {\bf
282} (13) 1987.

\bibitem{KM}
E. Kiritsis, N.A. Obers, B. Pioline, "Heterotic/Type II Triality and Instantons on $K_3$", JHEP {\bf 0001} (029) 2000,
arXiv:hep-th/0001083;

P. Mayr, S. Stieberger, "Threshold Corrections to Gauge Couplings in Orbifold Compactifications", Nucl. Phys.  B {\bf 407}
(725) 1993, arXiv:hep-th/9303017.

\bibitem{CPKW}
M. Cvetic, I. Papadimitriou, "Conformal Field Theory Couplings for Intersecting D-branes on Orientifolds", Phys. Rev. D
{\bf 68} (046001) 2003; Erratum-ibid. D70 (2004) 029903, arXiv:hep-th/0303083;

I. R. Klebanov and E. Witten, "Proton Decay in Intersecting D-brane Models", Nucl. Phys. B {\bf 664} (3) 2003, arXiv:hep-
th/0304079;

\bibitem{GMB}
E.Gava, J.F.Morales, K.S.Narain, G.Thompson, "Bound States of Type I D-Strings", Nucl.Phys. B {\bf 528} (95) 1998,
arXiv:hep-th/9801128;

Massimo Bianchi, "Type I Superstrings without D-branes", arXiv:hep-th/9702098;

C. Bachas, "(Half) a Lecture on D-branes", arXiv:hep-th/9701019.

\bibitem{MBJFMFF}
M. Bianchi, F. Fucito, J. F. Morales, "Dynamical supersymmetry breaking from unoriented D-brane instantons", JHEP {\bf
0908} (040) 2009, arXiv:0904.2156.


\end{thebibliography}
\end{document}